\documentclass{eas}
\usepackage{graphicx}
\begin{document}

\title{Precise Astrometry of Visual Binaries with Adaptive Optics.\\
       A way for finding exoplanets?}
\runningtitle{He\l miniak \& Konacki: Astrometry of Binaries with AO} 
\author{Krzysztof He{\l}miniak}\address{Nicolaus Copernicus Astronomical Center, 
Rabia\'nska 8, 87-100 Toru\'n, Poland}
\author{Maciej Konacki}\sameaddress{1}
\begin{abstract}
We present the results of our study of astrometric stability of
200-in Hale (Mt. Palomar) and 10-m Keck II (Mauna Kea) telescopes, both with Adaptive
Optics (AO) facilities. A group of nearby visual binaries and multiples was observed in near infrared, relative separations and position angles measured. We have
also checked the influence of some systematic effects (e.g. atmospherical refraction,
varying plate scale factor) on result and precision of astrometric measurements. We 
conclude that in visual binaries astrometrical observations it is possible to achieve
much better precision than 1 miliarcsecond [$mas$], which in many cases allows 
detection of the astrometrical signal produced by planetary-mass object.
\end{abstract}


%
\maketitle
\section{Introduction}
Astrometry is thought to be the most promising method of exoplanets detection in the
future. On the contrary to radial velocities measurements (RV), astrometry is almost 
independent on stellar physics, e.g. phenomena like the activity or pulsations, rather than the distance to the object. Todays interferometers are able to achieve precision 
at the level of mili- or microarcseconds, which is good enough to attempt exoplanet
research.

Nevertheless, the same level is possible to achieve in small fields by CCD imaging 
with adaptive optics (AO) systems. To do that, one must subtract and correct some 
systematical effects in order to obtain a gaussian distribution of the measurements.
In such case, the precision improves like $N^{-0.5}$, where $N$ is a number of single 
measurements. The goal of our studies was to check if a gaussian statistic can really 
be achieved with two top-class telescopes with AO systems: the 200-in Hale telescope + 
PHARO camera (Mt. Palomar) and 10-m Keck II telescope + NIRC2 (Mauna Kea), both working
in infrared (IR). With these facilities we were imaging 17 objects, majority of which 
were M-type dwarf binaries, located closer than 20 $pc$ from the Sun.

\section{Observations}
Most of the observations were made with Hale telescope and {\it Palomar High-Angular 
Resolution Observer} (PHARO). The camera was using three wide-bandpass filters -- K, 
K' and Ks -- and two narrow-band, centered on Br$\gamma$ and FeII lines (Hayward \etal
\ \cite{hay01}). PHARO's imaging mode was used with 39.91 and (mostly) 25.10 $mas/pix$ 
scale. With this telescope we made about 30,000 images of 12 binaries/multiples -- 
these are: GJ 195, GJ 352, GJ 458, GJ 507, GJ 661, GJ 767, GJ 860, GJ 873, GJ 9071 
and MWC 1080 
-- and two fields in open clusters, i.e. NGC 1039 and NGC 6871. Observations were made 
between October 2001 and November 2002, during 11 nights in groups of 2 to 4. Objects 
were observed in $dithering$ mode, which means small shifts of the telescope's 
position in order to subtract a variable IR background and reduce the influence of 
the chip's distortion.

Three other multiples were observed with Keck II + {\it Near Infra-Red Camera 2}
(NIRC2). These were 56 Per, GJ 300 and GJ 596. 
Main components were saturated, so the astrometry could be done only for double 
secondaries\footnote{Two ''secondary'' stars of GJ 300 are actually field stars. The 
same is for GJ 873.}. Keck II was used 4th March 2002 in imaging mode with 9.942 and 
39.686 $mas/pix$ scale, J, K' and K-cont (narrow-band) filters, also with {\it 
dithering} and, additionally, field rotation. 

\section{The method}
The closure of companions allows one to observe visual binaries 
in a smaller field, where atmospheric distortions are
weakly affecting the measurements, star's image is better sampled, and the 
measurements errors in pixels transfer to smaller errors in arcseconds. But, instead of parallax and proper motion, one have to take into account the orbital motion. Note also that in case of a positive detection of a 3-rd body, relative astrometry does not
give the information around which particular star of the system the body is orbiting. There are also some systematic effects, which, in every astrometric 
research, need to be corrected in order to achieve gaussian statistics (white noise)
of the measurements. For checking if the white-noise-behavior is present, we used
Allan variance ($\sigma_A$), which for a random dispersed values shows -1 slope on a 
log-log plot of $\sigma_A$ vs. so-called {\it lag}\footnote{For more details see 
Pravdo \& Shaklan (\cite{pra96}) or Lane \& Muterspaugh (\cite{lan04}).}.

\subsection{AO correction}
Proper AO correction is a crucial factor. Our example of GJ 352 (separation $\simeq
350\, mas$) shows how the quality of measurements is dependent on it. For only 10 properly AO-corrected frames, $rms$ of the separation $\rho$ was 4.2 $mas$ while for 62
others (in some of them the components were not resolved), $rms$ reached 19.8 $mas$.  

\subsection{PHARO chip geometry}
Depending on the position of the binary on the chip, we were obtaining various values
of $\rho$, what obviously reflects variations of the plate-scale around the chip. We adopted a model of distortion with a linear dependency of the plate-scale on the $x$ and $y$ coordinates of the chip, similar to the one found by Metchev \& Hillenbrand 
(\cite{met04}). After this correction, the measurements truly followed the 
white-noise-behavior (Fig. 1).

We have also found a small rotation of PHARO, between August and November 2002, by a value of $0.641 \pm 0.013\, deg$. This is also in agreement with results obtained by
Metchev \& Hillenbrand (\cite{met04}).

\subsection{Atmospheric refraction}
Precision bellow 1 $mas$ requires atmospheric refraction to be carefully taken into 
account. It's zenithal distance ($z$) dependency causes that observed separation of 
two stars is smaller than the true one (due to stars' different $z$'s). In our 
research we used a semi-full approach, described by He\l miniak (\cite{hel08}), which 
is suitable for IR.

\section{Precision and detection limits}
Assuming $3\sigma$ detection limit and transferring Pravdo \& Shaklan's (\cite{pra96})
Equation 8 (Sect. 6, p. 270), we get a relation:
\begin{equation}
a M_P [AU \cdot M_{Jup}] = 1562.5 \frac{\sigma_{\rho} M_S}{\pi} \left[ \frac{mas \cdot M_{\odot}}{mas} \right]
\end{equation}
where $a$ is a semi-major axis of a planet, $M_P$ is it's mass, $\sigma_{\rho}$ is the 
precision of separation measurements, $M_S$ is the star's mass and $\pi$ is it's 
parallax. $3\sigma_{\rho}$ is used instead of the astrometric signal $\Theta$. This 
relation was used to calculate, what kind of planets can be found in a
particular system. For stars from our sample with known mass and distance, best 
precision achieved and it's corresponding limits are given in Table 1. The values show
that with Hale and Keck II telescopes it is possible to find Jupiter- or even 
Saturn-mass bodies around given stars.

For several systems, observed in all epochs, we also calculated a level of a long-term
astrometric precision. We fitted 2-nd order polynomials, and also obtained $rms$' at 
the level of 100 $\mu as$ or sometimes better\footnote{Measurements, achieved precisions, limits and fittings for all systems can be found at http://www.ncac.torun.pl/$\sim$xysiek/source/mgr.pdf Ch.6, pp.94-105 (in polish, uncorrected for distortion!).}.

\section{Conclusions}
Todays telescopes with AO systems allows us to perform astrometric measurements with
precision well bellow 1 $mas$. This refers to a single-epoch observations, as well as
to a long-term stability. For many cases such a precision means an ability to detect 
exoplanets around nearby stars.

\begin{figure}
\includegraphics[width=0.95\columnwidth]{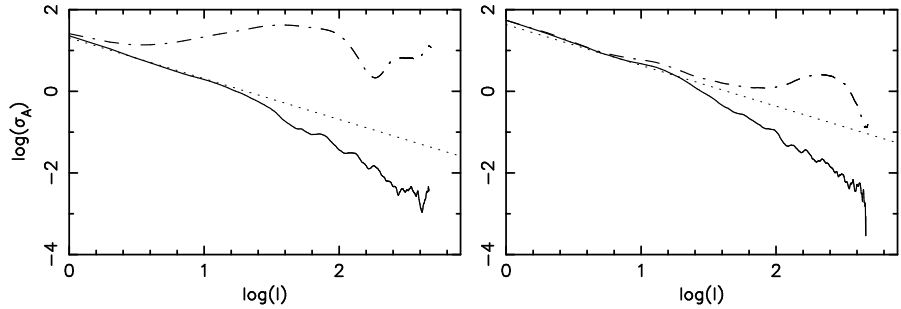}
\caption{Example of Allan variance vs. lag for uncorrected (dash-dotted) and corrected (solid line) measurements of separation components $x$ (left) and $y$. A small decrease is normal for finite series of measurements. Dotted line shows a --1-behavior for an ideal, infinitely long white-noise series.}
\end{figure}

\begin{table}
\caption{Smallest $\sigma_{\rho}$ and corresponding detection limits for researched stars.}
\begin{tabular}{c|c|c|cc|cc|c}
\hline
Star & Lowest & Dist. & Mass A & Limit for A & Mass B & Limit for B& Tel. \\
(GJ No.)& $\sigma\, [mas]$ & $[pc]$ & $[M_{\odot}]$ & $[AU \cdot M_J]$ & $[M_{\odot}]$ & 
$[AU \cdot M_J]$ &\\
\hline
195  & 0.12  & 13.89 & 0.53  &  1.38 &  0.19 &  0.50 & Hale \\
352  & 1.11  & 10.53 & 0.44  &  8.04 &  0.41 &  7.49 & Hale \\
458  & 0.28  & 15.32 & 0.40  &  2.68 &  0.37 &  2.48 & Hale \\
507  & 0.33  & 13.16 & 0.46  &  3.12 &  0.37 &  2.51 & Hale \\
569B & 0.11  &  9.81 & 0.071 & 0.116 & 0.054 & 0.088 & Keck \\
661  & 0.038 &  6.32 & 0.379 &  0.16 &  0.34 &  0.15 & Hale \\
767  & 0.09  & 13.35 & 0.44  &  0.83 &  0.40 &  0.75 & Hale \\
860  & 0.048 &  4.01 & 0.34  &  0.10 &  0.27 &  0.09 & Hale \\
873  & 0.57  &  5.05 & 0.36  &  1.62 & unknown & unknown & Hale \\
9071 & 0.20  & 13.89 & 0.53  &  2.22 &  0.49 &  2.05 & Hale \\
\hline 
\end{tabular}
\end{table}


\end{document}